\documentstyle[natb_209]{mn} 

\def\dosingle#1::::{#1}  \def\dodouble#1::::{ } 

\dodouble \documentstyle[natb_209,doublespacing]{mn} ::::

\def\nice#1::::{#1}    \def\subm#1::::{}   

\newcommand\docircappendix[1]{}

\newcommand\zzz[2]{#2}  

\def\SS{Sect.~}

\def\apj{ApJ}                 
\def\aap{A\&A}            
\def\mnras{MNRAS}
\def\araa{AnnRevA\&A}

\nice \input epsf ::::
\nice \def\mycaptionfont{\protect\footnotesize} ::::

\renewcommand\citep[1]{(\citealt{#1})}
\newcommand\citepf[1]{(\citealt*{#1})}    

\def\centreline{\centerline}
\def\.{{\cdot}} 
\def\gtapprox{\,\lower.6ex\hbox{$\buildrel >\over \sim$} \, }
\def\ltapprox{\,\lower.6ex\hbox{$\buildrel <\over \sim$} \, }
\def\propapprox{\,\lower.6ex\hbox{$\buildrel \propto\over \sim$} \, }

\def\arcs{\ifmmode {'' }\else $'' $\fi}     
\def\arcm{\ifmmode {' }\else $' $\fi}       
\def\deg{\ifmmode^\circ\else$^\circ$\fi}    

\def\fr7{7$ \hskip -0.9ex \vrule height0.8ex width0.8ex depth-0.73ex
                                                                \hskip0.1ex$}
\def\frtoday{Le\space\number\day\space\ifcase\month\or
  janvier\or f\'evrier\or mars\or avril\or mai\or juin\or
  juillet\or ao\^ut\or septembre\or octobre\or novembre\or d\'ecembre\fi\space \number\year}

\newcommand\joref[5]{#1, #5, {#2, }{#3, } #4}  
 
\newcommand\epref[3]{#1, #3, #2}

\def\cqg{ClassQuantGra}   %

\def\hMpc{\mbox{h$^{-1}$ Mpc}}

\def\rinj{{r}_{\mbox{\rm \small inj}}}  
\def\rSLS{r_{\mbox{\rm \small SLS}}}  
\def\Sdopp{S_{\mbox{\rm \small Dopp}}}
\def\Smax{S_{\mbox{\rm \small max}}}

\def\nhati{{\bf  {\hat{n}}_i}}    
\def\nhatj{{\bf {\hat{n}}_j}}    
\def\exunit{{\bf e_x}}  
\def\eyunit{{\bf e_y}}  
\def\ezunit{{\bf e_z}}  

\begin{document}

\title[COBE and Global Topology: Identified Circles]{COBE and 
Global Topology: An Example of the 
Application of the Identified Circles Principle}
\author[B.~F.~Roukema]{Boudewijn F. Roukema\\
Inter-University Centre for Astronomy and Astrophysics 
Post Bag 4, Ganeshkhind, Pune, 411 007, India\\ Email: boud@iucaa.ernet.in}
\def\today{\frtoday}

\maketitle


\begin{abstract}
The significance to which the cosmic microwave background
(CMB) observations by the satellite COBE can be used to
refute a specific observationally based hypothesis for the 
global topology (3-manifold) of the Universe is investigated,
by a new method of applying the principle of matched circle pairs. 

Moreover, it is shown that this can be done without assuming Gaussian
distributions for the density perturbation spectrum.

The Universe is assumed to correspond to a flat 
Friedmann-Lema\^{\i}tre model with
a zero value of the cosmological constant. 
The 3-manifold is hypothesised to be a 2-torus in two
directions, with a third axis larger than the horizon 
diameter. The positions and lengths of the 
axes are determined by the relative positions of the
galaxy clusters Coma, RX~J1347.5-1145 and CL~09104+4109, 
assumed to be multiple topological images of a single, physical
cluster.

If the following two assumptions are valid: 
(i) that the error estimates in the COBE DMR data are accurate 
estimates of the total random plus systematic error; and
(ii) that the temperature fluctuations are dominated 
by the na\"{\i}ve Sachs-Wolfe effect; 
then 
the distribution of the temperature differences between
multiply imaged pixels is significantly wider than the
uncertainty in the differences, 
and the candidate is rejected at the 94\%
level.

\dodouble \end{abstract} \clearpage
 \begin{abstract}  ::::  

This result is valid for either the `subtracted' or 
`combined' Analysed Science Data Sets, for either
$10\deg$ or $20\deg$ smoothing, and is slightly strengthened
if suspected contaminated regions from the galactic centre
and the Ophiuchus and Orion complexes are removed.

\end{abstract}


\dodouble \clearpage :::: 


\def\tcluster{
\begin{table}
\caption{The positions of the galaxy clusters 
Coma, CL~09104+4109 and RX~J1347.5-1145 shown as
J2000 right ascension ($\alpha$) and declination ($\delta$) in decimal degrees 
and redshift $z$. [Note: the label `J2000' in Fig.~1 of 
\protect\citet{FabCr95} is interpreted as a typographical error 
for `B1950'.]
\label{t-cluster}}
$$\begin{array}{l c c c} \hline 
 & \alpha & \delta & z  \\ \hline
\mbox{Coma} & 12.9969 &  27.9807 & 0.0239\\
\mbox{CL~09104+4109} & 9.2289 &  40.9428 & 0.442 \\
\mbox{RX~J1347.5-1145} & 13.7918 & -11.7533 & 0.451 \\
\hline
\end{array}$$
\end{table}
}  

\def\tasdswght{
\begin{table}
\caption{Weights for the three frequencies of DMR measurements
used for the `subtracted' and `combined' maps 
\protect\citep{Bennett92,Bennett94}.
\label{t-asdswght}}
$$\begin{array}{l c c c} \hline 
 & 31 \mbox{\rm GHz} & 53 \mbox{\rm GHz} & 90 \mbox{\rm GHz} \\ \hline
\mbox{subtracted} & -0.34 & 0.82 & 0.70  \\ 
\mbox{combined} & -0.49 & 1.42 & 0.18 \\
\hline
\end{array}$$
\end{table}
}  

\def\tresults{
\begin{table}
\caption{The values of the statistics $d,$ $\sigma,$
$S$ and $\Sdopp$, as defined in eqs~(\protect\ref{e-dmean}), 
(\protect\ref{e-sigma}), (\protect\ref{e-corr}) and 
(\protect\ref{e-cdopp}), for the COBE data.
\label{t-results}}
$$\begin{array}{ c c c c}\hline
d & \sigma & S & \Sdopp  \\ \hline
\multicolumn{4}{l}{\mbox{`Subtracted' Analysed Science Data Set: }}\\
\multicolumn{4}{l}{\mbox{$10\deg$ FWHM smoothing}} \\
-0.04 & 2.06 & 0.00 & -0.01 \\
\multicolumn{4}{l}{\mbox{$20\deg$ FWHM smoothing}} \\
-0.20 & 2.71 & -0.02 & -0.01 \\
\multicolumn{4}{l}{\mbox{`Combined' Analysed Science Data Set:}}\\
\multicolumn{4}{l}{\mbox{ 10$\deg$ FWHM smoothing }} \\
-0.01 & 1.54 & 0.02 & -0.01 \\
\multicolumn{4}{l}{\mbox{ 20$\deg$ FWHM smoothing }} \\
 -0.14 & 1.93 & 0.02 & -0.01 \\
\multicolumn{4}{l}{\mbox{20$\deg$ FWHM; excluding G.C., Orion, Oph loop}} \\
-0.02 & 2.05 & -0.03 & -0.03 \\
\hline
\end{array}$$
\end{table}
}  

\def\fhow{
\begin{figure}
\centering 
\nice \centreline{\epsfxsize=8cm
\zzz{\epsfbox[0 0 571 515]{"`gunzip -c howcircle.ps.gz"} }
{\epsfbox[0 0 571 515]{"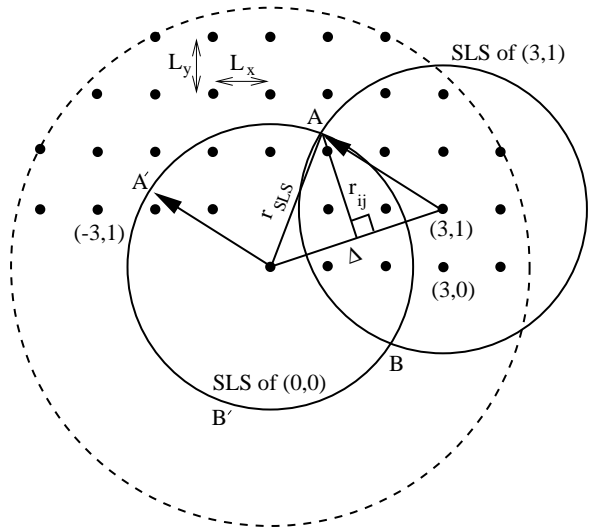"}}  } ::::
\caption[]{ \mycaptionfont
Geometry in covering space showing how an observer sees multiple
copies of circles on the CMB. 
Multiple copies of the observer which need to be
considered are shown as black dots, and can labelled
$(i,j)$ in units of the size of the Universe in 
the two small, fundamental directions. For clarity, 
the size of the Universe relative to the SLS (surface of
last scattering) is larger than 
that of the candidate under consideration in this paper.
The copy of the observer at $(0,1)$ is omitted for clarity.
An image of the observer, for example, at $(3,1),$ is separated
from the observer by a distance $\Delta$. The intersection of
the SLS spheres of the two observers is a circle, of which two
points lie in the plane of the figure. One such point is at A.
Because
the two observers are equivalent, the circle as seen by the
observer at $(3,1)$ is equivalent to a circle seen in the
same direction by the observer at the origin. 
That is, observer $(3,1)$ looking towards A is equivalent to
observer $(0,0)$ looking towards A$'$, and similarly for 
B and B$'$. These two {\em multiply imaged} points in the plane
are part of the the multiply imaged circle in three dimensions.
The radius of the identified circles
is $r_{ij}$. No identified circle is a great circle [except if
$(0,0)$ were counted as a multiple image of the observer, which would
not be useful]. The SLS has radius $\rSLS.$ The dashed circle of
radius $2\rSLS$ is the maximum extent to which multiple images
of the observer need to be considered.
}
\label{f-howcircle}
\end{figure}
} 

\def\fcircles{ 
\begin{figure*}
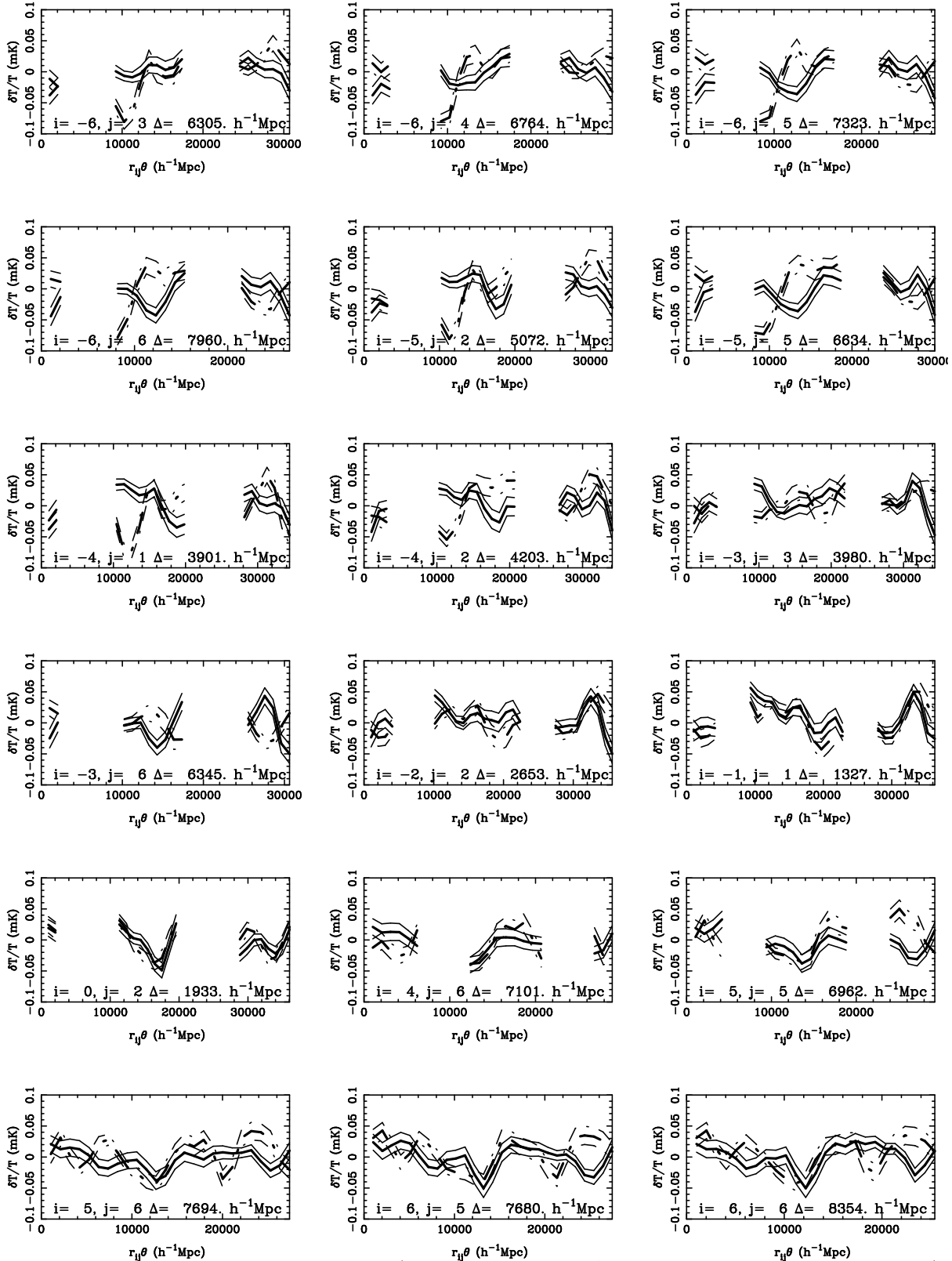

\centering 
\nice \centreline{\epsfxsize=16cm
\zzz{\epsfbox[ 40 42 587 761]{"`gunzip -c circles.ps.gz"}}
{\epsfbox[ 40 42 587 761]{"circles.ps"}}  } ::::
\caption[]{ \mycaptionfont
Temperature fluctuations in COBE DMR data 
(the `combined' 
four year Analysed Science Data Set, smoothed by $20\deg$)
around a selection of hypothetically identified
circles in the CMB, modelled in the covering
space, for $\Omega_0=1,$ $\lambda_0=0$, shown against the distance
around each circle. Thick lines show the values $\delta T/T$ 
and thin lines show these values plus or minus the $1\sigma$ uncertainties.
The horizontal length is the circle circumference. 
(By definition, no circles are great circles.)
Each pair of circles is defined by 
the plane halfway between the observer 
and a topological image of the observer at
$i L_x \exunit + j L_y \eyunit$.
The indexes $i$ and $j$ label each plot. 
The solid curves are for a circle at $(i,j)$ and the dashed curves
for its matched circle at $(-i,-j).$
For the torus family of flat 3-manifolds, as in this case, 
the circles of a 
pair are parallel. The distance between the two members of
a pair is therefore defined and is 
indicated here by $\Delta.$ Galactic latitudes between
$-20\deg$ and $20\deg$ are excluded from the plots. 
}
\label{f-circles}
\end{figure*} 

} 

\def\fcirclesall{
\begin{figure*}
\centering 
\nice \centreline{\epsfxsize=16cm
\zzz{\epsfbox[40 28 587 761]{"`gunzip -c circlesA.ps.gz"}} 
{\epsfbox[40 28 587 761]{"circlesA.ps"}}  } ::::
\caption[]{ \mycaptionfont
Four pages showing full set of temperature fluctuations, apart from those
which have no useful points, due to the galactic cut and/or circle
radii which are smaller than the COBE resolution, are not
shown. These are included as an appendix in the electronic version 
only. Figure~\protect\ref{f-circles} constitutes a subset of
Figs~\protect\ref{f-circlesA} - \protect\ref{f-circlesD}. 
}
\label{f-circlesA}
\end{figure*}

\begin{figure*}
\centering 
\nice \centreline{\epsfxsize=16cm
\zzz{\epsfbox[40 28 587 761]{"`gunzip -c circlesB.ps.gz"}} 
{\epsfbox[40 28 587 761]{"circlesB.ps"}}  } ::::
\caption[]{ \mycaptionfont
Temperature fluctuations around circles in the COBE
data (continued).
}
\label{f-circlesB}
\end{figure*}

\begin{figure*}
\centering 
\nice \centreline{\epsfxsize=16cm
\zzz{\epsfbox[40 28 587 761]{"`gunzip -c circlesC.ps.gz"}} 
{\epsfbox[40 28 587 761]{"circlesC.ps"}}  } ::::
\caption[]{ \mycaptionfont
Temperature fluctuations around circles in the COBE
data (continued).
}
\label{f-circlesC}
\end{figure*}

\begin{figure*}
\centering 
\nice \centreline{\epsfxsize=16cm
\zzz{\epsfbox[40 406 587 761]{"`gunzip -c circlesD.ps.gz"}} 
{\epsfbox[40 406 587 761]{"circlesD.ps"}}  } ::::
\caption[]{ \mycaptionfont
Temperature fluctuations around circles in the COBE
data (continued).
}
\label{f-circlesD}
\end{figure*}

} 

\def\faitoff{ 
\begin{figure*}
\centering 
\nice \centreline{\epsfxsize=15cm
\zzz{\epsfbox[10 -13 731 378]{"`gunzip -c aitoff.ps.gz"}} 
{\epsfbox[10 -13 731 378]{"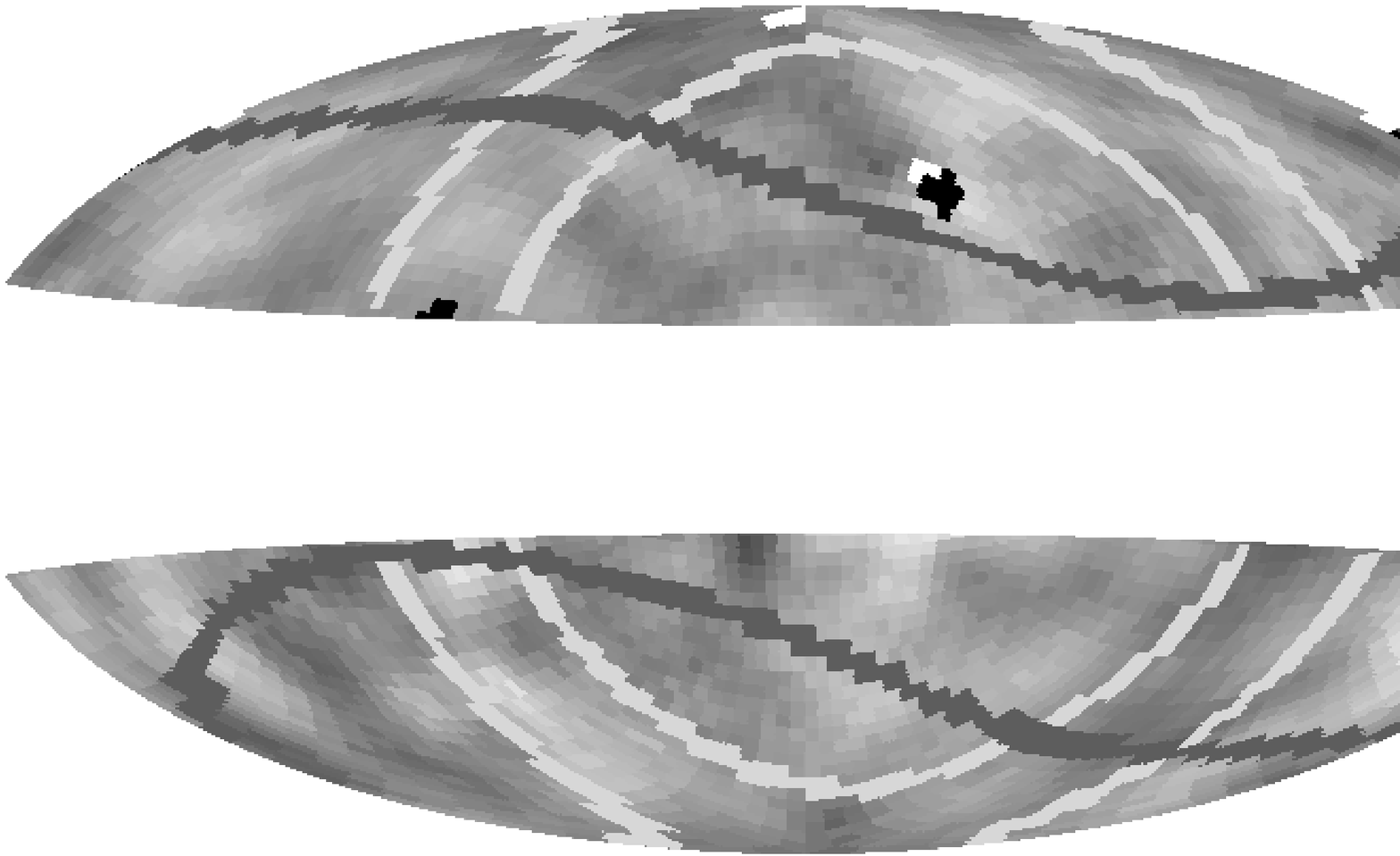"}}  } ::::
\caption[]{ \mycaptionfont
COBE DMR 10$\deg$ smoothed `combined' ASDS map in Aitoff projection, 
with galactic latitude $b^{II}$ increasing from $-90\deg$ to $90\deg$ from
bottom to top and galactic longitude $l^{II}$ 
decreasing from $180\deg$ at the left
to $-180\deg$ at the right. A galactic cut of $\pm20\deg$ is
applied. The positions of the three 
hypothesised fundamental axes determined by the three cluster
images are indicated by 
black
`spots'. From right to left, these 
are at $(-179\deg,41\deg)$, 
determined by the vector from Coma to CL~09104+4109, 
$(-35\deg,46\deg),$
the vector from Coma to RX~J1347.5-1145, and 
at $(75\deg,18\deg)$ the  
orthogonal third axis. 
The positions of the clusters (Sun centred observer) are shown 
by white `spots': Coma is near the north galactic pole, the other two 
clusters are 
$\sim 3\deg$ north of the corresponding axis directions.
{\em Identified circles} for $(i,j)=(-2,2)$ and  $(i,j)=(5,6)$ 
are shown as light and dark grey curves respectively
(cf Fig.~\protect\ref{f-circles}).
\label{f-aitoff}}
\end{figure*} } 

\def\fslong{ 
\begin{figure}
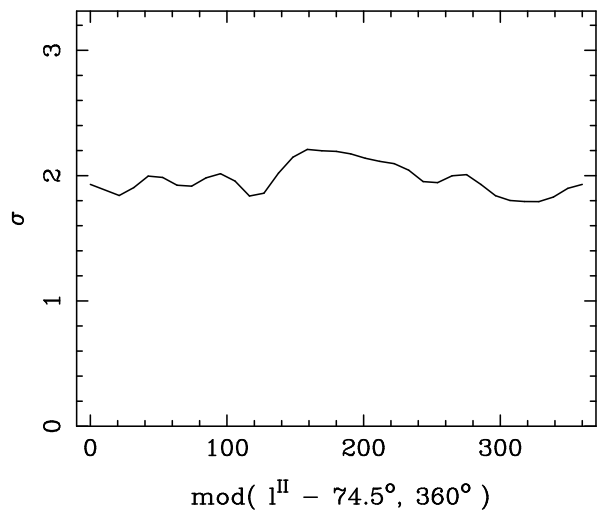

\centering 
\nice \centreline{\epsfxsize=8cm
\zzz{\epsfbox[42 31 459 380]{"`gunzip -c circ_long_s.ps.gz"}} 
{\epsfbox[42 31 459 380]{"circ_long_s.ps"}}  } ::::
\caption[]{ \mycaptionfont
Dependence of $\sigma$, i.e. the standard deviation from identity 
of temperature fluctuations around
hypothetically multiply imaged circles on the CMB, in units
of the observational uncertainties
[eq.~(\protect\ref{e-sigma})]. The value at galactic longitude
$l^{II}=74.5\deg$ is for the \protect\citet{RE97} $T^2$ 
cluster-based hypothesis and is clearly greater than $\sigma=1.$
Points at other longitudes are for 
other hypotheses equivalent to that of 
\protect\citeauthor{RE97}, except for a rotation in galactic longitude.
The latter are unmotivated by observations.
This and
the following three figures are for the $20\deg$ smoothed `combined'
map.}
\label{f-slong}
\end{figure} } 

\def\fdlong{ 
\begin{figure}
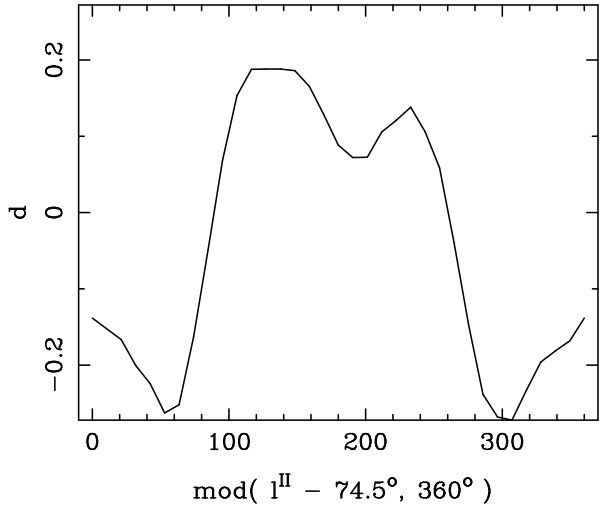

\centering 
\nice \centreline{\epsfxsize=8cm
\zzz{\epsfbox[42 31 459 380]{"`gunzip -c circ_long_d.ps.gz"}} 
{\epsfbox[42 31 459 380]{"circ_long_d.ps"}}  } ::::
\caption[]{ \mycaptionfont
The mean temperature difference $d$ [eq.~(\ref{e-dmean})] 
along identified circles is
plotted, as for Fig.~\protect\ref{f-slong}.
\label{f-dlong}}
\end{figure} } 

\def\fclong{ 
\begin{figure}
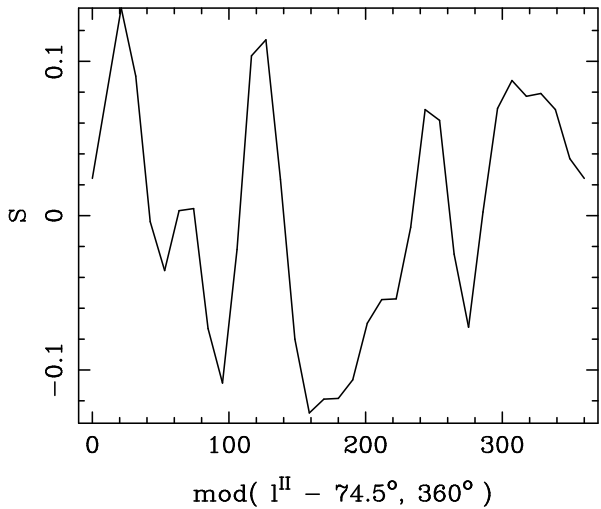

\centering 
\nice \centreline{\epsfxsize=8cm
\zzz{\epsfbox[42 31 459 380]{"`gunzip -c circ_long_c.ps.gz"}} 
{\epsfbox[42 31 459 380]{"circ_long_c.ps"}}  } ::::
\caption[]{ \mycaptionfont
Dependence of $S$ [eq.~(\ref{e-corr})], the correlation statistic defined by 
\protect\citet{Corn98b}, along identified circles as for 
Fig.~\protect\ref{f-slong}.
\label{f-clong}}
\end{figure} } 

\def\fplong{ 
\begin{figure}
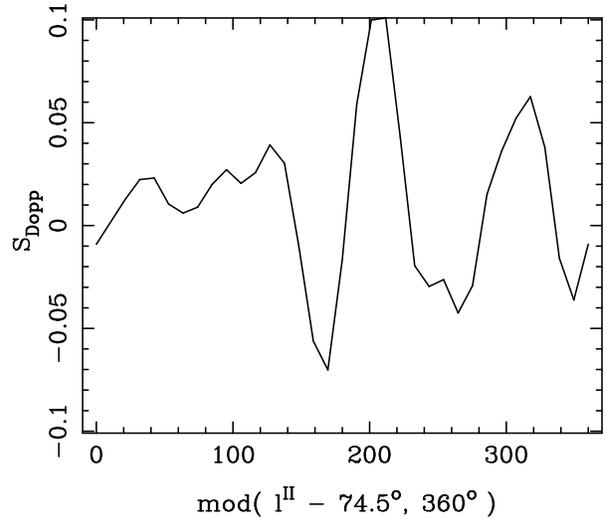

\centering 
\nice \centreline{\epsfxsize=8cm
\zzz{\epsfbox[42 31 459 380]{"`gunzip -c circ_long_p.ps.gz"}} 
{\epsfbox[42 31 459 380]{"circ_long_p.ps"}}  } ::::
\caption[]{ \mycaptionfont
Dependence of $\Sdopp$
 [eq.~(\ref{e-cdopp})], a modification of the $S$ statistic 
suggested for detecting correlations of the Doppler effect,
along identified circles as for 
Fig.~\protect\ref{f-slong}.
\label{f-plong}}
\end{figure} } 


\section{INTRODUCTION}

\subsection{Friedmann-Lema\^{\i}tre models: curvature and topology}

At the end of last century, \citet{Schw00,Schw98} 
dared to suggest `a few remarks' which require `a total break
with the astronomers' deeply entrenched views'.
He suggested (i) that geometers' abstract concepts of non-Euclidean geometries
could correspond to the real Universe, but also (ii) that 
non-trivial topology could also be required in order to represent
the space in which we live. 
Since then, the physical theory (general relativity) 
linking his first suggestion, curvature, to the stress-energy tensor
has encouraged considerable observational efforts in measuring
(directly or indirectly) the curvature of the observable Universe.

On the other hand, 
mathematical difficulties and the lack of an accepted quantum gravity
theory probably explain why less observational effort has 
been directed at following up the second of
Schwarzschild's two remarks that have
`neither any real practical applicability nor any pertinent 
mathematical meaning', i.e. 
measuring the topology of the Universe. Since 
the curvature estimates strongly favour a non-positive curvature, 
without measuring the topology of the Universe it is not possible
to answer fundamental questions, such as whether or not 
the Universe is finite. (Measuring topology does not guarantee
an answer, but not measuring it guarantees not having an answer, except
if the curvature is positive.)

Both curvature and topology have been put in a relativistic context
in the Friedmann-Lema\^{\i}tre models \citep{deSitt17,Fried24,Lemait58}.

Mathematical developments in three-dimensional geometry and topology 
\citep{Thur82,Thur97} now abound, and 
theoretical ideas of the generation of topology at the
quantum epoch are developing rapidly
\citep{MadS97,Carl98,Ion98,DowS98,DowG98,Rosal98,eCF98,Gibb98,RatT98}.

 Recently, considerable
work in observational methods, constraints and candidates for 
the global topology of the Universe has also commenced. 
These 
include (i) cosmic microwave background (CMB), i.e. essentially 
two-dimensional,  
methods (see \SS\ref{s-twodim})
 and (ii) three-dimensional methods 
\citep{Fag87,Fag96,LLL96,Rouk96,RE97,Fag98,LLU98,RB98}.

Work has also been done considering how simulations of
large scale structure (the galaxy distribution)
are affected by the usual assumption 
(for numerical rather than physical reasons) that the
Universe is multiply connected \citep{FMel}.

For theoretical reasons, interest
is probably highest in the hyperbolic manifolds. 
However, the observational arguments tending towards a flat, 
cosmological constant dominated universe 
\citepf{FYTY90,FortMD97,ChY97,SCP9812} may bring interest back to
more thorough analyses of flat, multiply connected
Friedmann-Lema\^{\i}tre models. Moreover, (i) 
for a cosmological constant to be observable, fine-tuning
is necessary if an inflationary scenario is accepted, 
which may also fine-tune the topology scales
such that $\rinj < \rSLS$\footnote{The 
shortest distance between two topological images of an 
astrophysical object is termed the injectivity diameter, $2\rinj$,
and can be thought of the `smallest size' of a small universe. 
The distance to the surface of last scattering (hereafter, SLS) is
labelled $\rSLS$.};
and (ii) the unknown physics required to explain 
$\lambda_0 > 0$ might also explain the topology of 
the Universe.\footnote{The dimensionless cosmological constant
is defined $\lambda_0 \equiv c^2 \Lambda/(3 H_0^2)$; in the
notation of \protect\citet{Peeb93}, 
$\lambda_0 \equiv \Omega_\Lambda $.}

Hence, the flat models should not be neglected.

For reviews on cosmological topology, see 
\citet{LaLu95,Lum98,Stark98} and \citet{LR99}.

The 2-D and 3-D methods have different underlying assumptions on
the physics of regions of 
space or of objects, so systematic errors are likely to 
differ between the two approaches. Both 
have advantages and disadvantages. For a list and discussion of both
2-D and 3-D methods, 
see Table~2 of \citet{LR99}. A discussion restricted to the 2-D
methods follows here.

\subsection{Two-dimensional (CMB) methods for testing global topology}
\label{s-twodim}

The 2-D methods 
cited above either (i) model (analytically or by simulations) 
the perturbation field in the universal covering 
space\footnote{The universal 
covering space can be thought of
as an `apparent' universe composed of
multiple adjacent copies of the physical universe. It is a
useful, though unphysical, construct for calculations. The
physical universe itself (one `copy') is referred to 
as the fundamental polyhedron.} 
of compact universes and compare the
random realisations (or statistics) with CMB data 
\citep{Stev93,Sok93,Star93,Fang93,JFang94,deOliv95,dOSS96,LevSS98a,BPS98}
or (ii) analyse CMB data directly for multiple topological images
(see \SS\ref{s-matchedc}; \citealt{Corn96,Corn98b,Corn98a,Weeks98}). 
[The method of \citet{LevSS98b}
probably falls into the latter category, but is not discussed further.]

\subsubsection{Advantages and disadvantages 
in the perturbation simulation approach}  \label{s-disadvant}

The fact that the 
former method (i), 
(which has been applied principally to Euclidean models, 
i.e. toroidal or hexagonal prismical models\footnote{\protect\citet{BPS98}
applied this same method to two hyperbolic models, but using correlation
functions instead of eigenmodes. \protect\citet{Inoue99} has demonstrated
calculation of eigenmodes of hyperbolic models.})
makes statements about an ensemble of universe models can be 
interpreted as either an advantage or a disadvantage relative 
to the latter method (ii).
The simulational methods (a) test 
consistency of observational and model perturbation statistics,  
based on (b) statistical properties of primordial perturbations,
while the identified circles principle (\SS\ref{s-matchedc})
directly tests whether or not the SLS is consistent with 
multiple topological imaging, using the actual temperature 
perturbations measured.
\begin{list}{(\alph{enumi})}{\usecounter{enumi}}
 \item The problem with a result about perturbation statistics
is that the confidence level with which a
particular multiply connected model can be rejected is only 
a statement about the rarity of perturbation properties required
to be consistent with the observations.

On the contrary,
the circles principle enables rejection of a model by showing
that temperatures at supposedly multiply topologically imaged
points are inconsistent. 
That is, it can show that {\em no} set of perturbations
would allow the model to fit the data. 


\item In addition, 
both theoretical and observational
justification at large scales 
of the perturbation statistics assumed
for the simulational approach 
are generally made {\em under the assumption of simple connectedness}.
It could be considered 
inconsistent to test multiple connectedness if properties
expected under the assumption of simple connectedness were assumed.
\end{list}

The theoretical motivations for
Gaussian amplitude distributions and a 
$P(k) \propto k^1$ power spectrum are unlikely to be valid at scales
approaching $2\rinj$ and $r_+$. 

That is, either for a hyperbolic or for a
flat, $\lambda_0 \sim 0.7$ metric to be presently observable
(irrespective of topology),
inflation needs some degree
of fine-tuning [which can partly be provided by the ergodicity of
geodesics in the former case, \citet{Corn96}]. For a flat, 
$\Omega_0=1$ multiply connected model to be observable, even
more fine-tuning is needed.\footnote{$\Omega_0$ is the 
present value of the density parameter.}

So, either flat, observably multiply connected models can be 
rejected for purely theoretical reasons, or else they can be
tested observationally accepting that perturbation statistics 
may differ somewhat from inflationary expectations.

Observational motivation for
Gaussian amplitude distributions and a 
$P(k) \propto k^1$ power spectrum at the $2\rinj$ and $r_+$ scales 
is equally lacking --- 
if one is testing a multiply connected universe, since 
the only observational justification of these properties
on large scales is that of COBE data analysed
{\em assuming simple-connectedness}. 
Moreover, although several 
authors find consistency with Gaussian statistics 
(e.g. \citealt{Kogut95,Kogut96c,CollGP96}), others 
find indications of non-Gaussianity 
\citepf{Ferr98,Pando98}, which are further discussed
in \citet{BrTeg99}.

Apart from the problems inherent to the perturbation simulation
approach, there are some factors which could be added to 
this approach but have not yet been dealt with in the work cited 
above dealing with flat
multiply connected models:

\begin{list}{(\alph{enumi})}{\usecounter{enumi}}
\addtocounter{enumi}{2}
	\item Only three of these eight papers
discuss $T^2$ models, and only one compares such models
with the COBE observations in any detail.

	\item The integrated Sachs-Wolfe effect (ISW) is
not considered in any of the flat model papers.
[Note that the ISW {\em is} discussed in \citet{BPS98} for the
hyperbolic case.]

	 \item The possible contribution of redshifts in the
SLS radiation due to peculiar velocities at
the SLS (hereafter, `the Doppler effect') is not considered
in any of these papers. This is expected to contribute less than the
na\"{\i}ve Sachs-Wolfe effect to temperature fluctuations $\delta T/T$
on COBE scales, under the assumption of simple\-{-}con\-nect\-ed\-ness. 
However, if this assumption is dropped and $\rinj < \rSLS$ is
being tested, it is difficult to be certain about 
properties of the 
velocity field on scales near $\rinj$ based on present quantum
cosmology theory. 

	A conservative observational approach would
be to consider statistics which take into account this {\em locally
anisotropic} radiation.

	\item There is also a more local observational problem 
with the work done so far for flat and hyperbolic models.
What is important for falsifying a small universe hypothesis
is the significance of
individual fluctuations, and whether or not these are multiply imaged.

The significance of individual fluctuations has been studied by
\citet{CaySm95}. This paper suggests that some of
the most significant hot spots ({\em above} the 20 degree galactic cut)
are galactic contaminants, from the Orion and Ophiuchus 
complexes. Removal of such non-axisymmetric contaminants has not yet
been discussed in the small universe literature.

\end{list}

\subsubsection{Advantages and disadvantages 
in the identified circles principle approach}

 Although tests based on the identified circles principle
(\SS\ref{s-matchedc}) avoid the problems (a) and (b) above, 
they share a weakness with the perturbation
simulation approach by being adversely affected by
noise, in particular from detector noise and from 
residual foreground contamination, which
can make genuinely matched circles look different.

 The ISW and Doppler effects will also provide some 
contributions to CMB maps. The former may be treated as foreground 
contamination. The latter will provide locally
anisotropic emission, which should be taken into account as 
a modification of \citeauthor{Corn96}'s identified circles principle 
(\SS\ref{s-dopp}).

The former is a weakness of applications of the identified circles
principle, shared by the simulation approach, for 
non-flat models and models with a non-zero cosmological constant. 
The ISW, due to time varying of the gravitational potential,
is likely to be strong if the Universe is hyperbolic,
particularly at low redshifts and at the large 
angular scales present in the COBE maps 
\citep{Inoue99}. In that case, temperature fluctuations at low redshifts
are mixed in with the signal from the SLS, so that an identified
circles analysis or a simulation analysis restricted to the SLS
cannot account for the full signal at COBE scales.

\subsubsection{Optimal comparison of multiply topologically imaged points}

In view of the above comments,
it is clear that for the testing of flat, small universe
models, 
there are several advantages if the identified circles principle 
\citep{Corn96,Corn98b,Corn98a,Weeks98}
can be used, 
since it {\em avoids} (a) making 
statistical statements about perturbations and (b) requiring
assumptions about the statistics of the 
perturbations,\footnote{The 
method of \protect\citet{LevSS98b} can probably also
avoid (a) and (b).} even though it {\em shares} the problem of 
the perturbation simulation approach in its dependence on the
statistics of detector noise and residual foregrounds.

There already exists a specific candidate for a small, 
flat universe which has been suggested by observations
\citep{RE97,RB98}. Because
the candidate is fixed in astronomical coordinates, this 
offers a good example on which to see if COBE data can be
used to reject a small universe hypothesis {\em by using
only the COBE data itself, without making theoretical 
hypotheses about the perturbation properties.} 

\subsection{Observational rejection of 
\protect\citeauthor{RE97}'s (1997) $T^2$ candidate}

The purpose of this paper is to see if this hypothesis, considered
as a null hypothesis, can be rejected purely on the basis of the COBE
observations, of about 
$\sim 300$ effectively independent pixels at which temperature fluctuations 
have been measured,\footnote{Since the DMR (Differential
Microwave Radiometer) is a differential instrument, the beam-width 
of $7\deg$ per antenna implies a width of $10\deg$ smoothing in the 
difference maps; a $20\deg$ cut for the galactic plane is adopted here; 
hence about 300 independent pixels.} without making assumptions
regarding the statistics of primordial density perturbations.

\subsubsection{The $T^2$ candidate}

The candidate $T^2$ 3-manifold \citep{RE97,RB98}, according to which two galaxy 
clusters at redshifts $ z\sim 0.4$ would be topological images of 
the Coma cluster, is presented in \SS\ref{s-t2}. 

Apart from the striking fact that these three cluster images form a near right
angle (in 3-D) and  that right-angled `toroidal' multiply connected 
universes are those which have been most extensively compared to 
observations, the relative simplicity of the hypothesis makes it
ideal for testing if observational methods of refutation can really
be made statistically watertight, in preparation for future candidate
3-manifolds, which are unlikely to be this simple.

Methods of refuting the supposed identity of the three clusters 
by their individual properties (positions and velocities of galaxies, 
overall masses, X-ray fluxes), by implied multiple imaging of 
large scale-structure, by predicted positions and redshifts
of further topological images of the cluster and by proposals of 
observations relating to sub-cluster structures are discussed 
at length in \SS4.2, \SS4.3 and \SS4.4 of \citet{RBa99}.

\subsubsection{How to apply the identified circles principle to 
reject the $T^2$ candidate}

The basic principle of any method used to constrain or detect
candidates for the 3-manifold is that if the hypothesis of
trivial topology is not adopted, then in some cases, 
photons can travel between
a single, physical object or region of space to the observer
by several different geodesics of (in general) widely differing
lengths. That is, the objects or regions of space 
are observed at widely differing positions and redshifts.

In the case of the CMB, since the redshift is nearly equal for all
points on the SLS by definition, only special subsets of angular 
positions on the SLS would be multiply imaged. 

\citet{Corn96,Corn98b} realised that these special subsets form 
a well defined and easily described set in the covering space:
the subsets are pairs of matched circles. The explanation for this is
subtle, but elegant, and is briefly described in \SS\ref{s-matchedc}. 

{\em If the radiation from the SLS is locally isotropic}, then
the observed temperature fluctuations observed around two circles
of such a pair should be identical, to within the observational
uncertainty.

If the values of corresponding points on the circles 
are inconsistent, then 
the hypothesis can be rejected with a confidence level
based on the estimated uncertainty per resolution element, or pixel.

The assumption of local isotropy seems reasonable for the COBE
data, due to the large smoothing length (effectively $10\deg$).
At this scale, the
effective temperature correlations should be determined uniquely by the 
na\"{\i}ve Sachs-Wolfe effect \citep{SW67}, i.e. by gravitational 
redshifts depending on the depths or heights of the potential wells. 

As long as the gravitational potentials are spherically symmetric,
i.e. if the equivalents of filaments and great walls on a linear 
scale do not exist at $z\sim 1100$, then this should be a fair
assumption. If filaments and walls did exist, then since temperatures 
in two different COBE `pixels' on corresponding circles represent
{\em averages over nearly two-dimensional spherical sections 
at different angles in the fundamental polyhedron},
some modelling of the three-dimensional two-point correlation
function would be needed in order 
to calculate the probability of these averages
being identical.

For this paper, the averaging over the different spherical sections,
each of diameter $\sim 1000${\hMpc} (for $\Omega=1$) and 
$10\deg$ smoothing, 
is assumed to result in identical values.

\subsubsection{How to treat anisotropic (Doppler) radiation when applying
the identified circles principle}

The amplitude of the Doppler effect at the SLS, which is
{\em locally anisotropic}, is assumed to be negligible relative to the
na\"{\i}ve Sachs-Wolfe effect. 
Since the first Doppler `peak' is on 
scales smaller than those resolved by the COBE DMR experiment, the
fractional contribution of a Doppler component to an identified 
circles test of COBE data is probably small.

However, if `linear' filaments and
walls existed at the recombination epoch, or if velocity fields
related to the $\rinj$ scale existed, then this might not be correct.

It should also be remembered that the {\em intrinsic} spatial
separations in a multiply connected universe can be much smaller
in the fundamental polyhedron 
than in the covering space, so coherent velocity fields on 
apparently large, but inherently small, scales may generate
a Doppler effect which would be unexpected under the 
simple connectedness assumption.

Although the na\"{\i}ve Sachs-Wolfe effect is assumed here to 
be dominant and to average out to locally isotropic values, 
an extension of \citeauthor{Corn98b}'s (1998b) statistic 
which should be sensitive to Doppler dominated temperature
fluctuations is suggested here, just in case the Doppler component 
is more important than expected.

If the Doppler effect dominates, then for 
pairs of multiply imaged pixels which also have nearly parallel 
lines-of-sight, the values should be nearly equal in absolute value,
with equal signs if the angle between them 
is close to $0\deg$ and opposite signs
if the angle is close to $180\deg$. Pixels for which the lines of
sight are roughly orthogonal provide no theory-free constraint 
on topology if the Doppler effect is dominant.

Hence, a correlation function weighted by the dot product of the
two line-of-sight vectors of each pixel pair should measure 
the degree to which the sets of points represent identical 
velocity induced temperature fluctuations. A dot product
weighted correlation statistic is presented [eq.~(\ref{e-cdopp})] 
and demonstrated below.

\subsection{Structure of paper and conventions}

In \SS\ref{s-method}, the hypothesised 3-manifold is briefly presented
(\SS\ref{s-t2}), the matched circles principle is explained 
and the statistics used to estimate the consistency of 
multiply imaged `pixels' on the CMB are defined (\SS\ref{s-matchedc}),
and  the COBE data are briefly introduced (\SS\ref{s-cobe}).
Results are presented in \SS\ref{s-results} and  
discussed in \SS\ref{s-disc}.
Conclusions are in \SS\ref{s-conc}.

The Hubble constant is parametrised here 
as $h\equiv H_0/100$km~s$^{-1}$~Mpc$^{-1}.$ Comoving coordinates are
used
(i.e. `proper distances', \citealt{Wein72}, equivalent to `conformal time'
if $c=1$). The metric assumed is flat, i.e., 
$\Omega_0 + \lambda_0 \equiv 1,$
since the candidate
3-manifold is $T^2 \times R.$ 
The case $\lambda_0=0$ is the main case discussed here.

\section{METHOD} \label{s-method}

\subsection{The $T^2$ candidate} \label{s-t2}

\citet{RE97} pointed out that massive clusters of galaxies, dominated
by their hot, X-ray emitting gas, should be relatively 
good `standard candles' for observational constraints on 
the topology of the Universe. This was, in a sense, an update of
the `classical' searches for multiple topological images 
of single objects \citep{Gott80}. 

Although not a goal of that work, it was noticed that among
a small number (seven) of very bright clusters used to illustrate the
principle, three of these form nearly a right angle (to within 
$\sim 2-3$\% depending on $\lambda_0$) with nearly equal (to 1\%) 
arm lengths. This happens to be just what would be expected 
if space is $T^3$ with two side lengths equal and the
third unknown, or if it is $T^2 \times R.$ 
An {\em a posteriori} 
estimate of the probability of this occurring by chance is
difficult to make in an unbiased way, so observational
refutation is more prudent than theoretical refutation. 

Moreover, this 3-manifold is very simple, relative to other
flat 3-manifolds or to the hyperbolic
3-manifolds [which are considered more likely for several 
theoretical reasons, e.g. \citet{Corn98a,eCF98}]. The implied 
positions and redshifts of 
multiple topological images of objects are therefore
relatively easy to calculate and 
observationally refutable predictions can be made.
A list of the various observational arguments for and against
the candidate, apart from the present study, is provided
in the discussion section of \citet{RBa99}.

\nice \tcluster ::::

The fundamental polyhedron (hereafter, FP) considered is therefore 
a nearly square, long prism. Two axes are defined 
as, $L_x \exunit,$ 
the vector from the Coma cluster to the cluster 
CL~09104+4109 and 
$L_y \eyunit,$ the vector from the Coma cluster to 
the cluster RX~J1347.5-1145. The third axis $L_z \ezunit$ 
is defined as
the cross product of these two vectors and assumed to 
be larger than the horizon diameter ($L_z > 2R_H$). 
(The vectors $\exunit,$ $\eyunit$ and $\ezunit$ are unit vectors.)
The positions 
and redshifts adopted for the clusters 
are shown in Table~\ref{t-cluster}.

The values of $L_x \approx L_y$ vary slightly from $\approx 960${\hMpc}
to $\approx 1230${\hMpc} for $0 \le \lambda_0 \le 0\.9.$
$L_z$ is larger than the horizon diameter.

\subsection{The Matched Circles Principle} \label{s-matchedc}

 \nice \fhow :::: 

The matched circles principle \citep{Corn98b} is illustrated
in Fig.~\ref{f-howcircle} and explained in this figure's caption.
Topological images of the observer in the lower half of 
the image are redundant (they give the same set of identified
circles as the upper half).
Depending on the value of the cosmological constant,
$\lambda_0,$ the ratio of $\rSLS$ to $L_x \approx L_y$ 
for the $T^2$ candidate under consideration 
is 
$6.0 \ltapprox \rSLS/L_x \ltapprox 13.2$ for $0 \le \lambda_0 \le 0\.9.$
This implies 
about three to six times as many circle pairs as shown
in Fig.~\ref{f-howcircle}. Since some circles fall within or mostly 
within $20\deg$ of the galactic plane, the number which can be
usefully compared is smaller than this by about a third.


\subsubsection{Locally Isotropic Radiation: the Sachs-Wolfe effect}

To see if the measured temperature fluctuations are consistent with the 
hypothesis of physical identity of the circles, 
the most straightforward approach is  
to test the hypothesis that temperatures on corresponding pixels
are equal to within observational error.

This hypothesis is tested here by considering the difference 
in the temperature fluctuations in two 
corresponding pixels on matched circles to be a random realisation
of a Gaussian distribution centred on zero with a width determined
by the uncertainties of the measurements in the two pixels.
By normalising the distribution for each pair to have a standard
deviation of unity, the full set of pairs of multiply imaged
pixels can be combined to form a large sample of a single
distribution of mean zero and standard deviation unity.

This standard deviation is 
$\sigma,$
defined
\begin{equation}
\sigma^2(r)\equiv \left<
         {  \left[ \left({\delta T \over T}\right)_i 
                 - \left({\delta T \over T}\right)_j \right]^2
     \over { \left[\Delta \left({\delta T \over T}\right)\right]_i^2 +
                  \left[\Delta \left({\delta T \over T}\right)\right]_j^2 } }
\right> .
\label{e-sigma}
\end{equation}
where  $\left({\delta T \over T}\right)_i$ and 
$ \left({\delta T \over T}\right)_j$ are the temperature fluctuations
in the two (hypothetically) multiple images of a single region of 
space-time. 

The assumption that the mean difference is zero is likely to
be a good approximation independently of whether or not the hypothesis
is correct, since the dipole components have been 
subtracted off the data and little correlation should exist
on such large scales if the assumption of simple-connectedness
is correct. Nevertheless, it is useful to check
the mean difference, which is defined
\begin{equation}
d\equiv \left<
         {  \left({\delta T \over T}\right)_i 
                 - \left({\delta T \over T}\right)_j 
     \over { 
\left\{ \left[\Delta \left({\delta T \over T}\right)\right]_i^2 +
      \left[\Delta \left({\delta T \over T}\right)\right]_j^2 
\right\}^{1/2}  
} } \right>.
\label{e-dmean}
\end{equation}

It is also useful to consider the 
statistic of \citet{Corn98b}, which is essentially 
a two-point autocorrelation function normalised by the variance per
pixel [eq.~(2) of \citeauthor{Corn98b}], though the interpretation
is slightly less straightforward. This statistic is
\begin{equation}
S\equiv { \left<
         { 2 \left({\delta T \over T}\right)_i 
                  \left({\delta T \over T}\right)_j }\right> 
     \over { 
\left< \left({\delta T \over T}\right)_i^2 +
\left({\delta T \over T}\right)_j^2  \right> } } .
\label{e-corr}
\end{equation}

\subsubsection{Doppler Effect} \label{s-dopp}

While it is possible that with the resolution of COBE, 
the Doppler effect averages out 
from three-dimensional space to 
the one-dimensional corresponding circles in a way which makes
it negligible, 
the following variation on eq.~(\ref{e-corr}) is proposed. 
This modified statistic is 
\begin{equation}
\Sdopp \equiv {
         { \left< 2 \left({\delta T \over T}\right)_i 
                  \left({\delta T \over T}\right)_j 
                   	\nhati {\boldmath .} \nhatj \right> }
     \over { 
\left<
\left({\delta T \over T}\right)_i^2 +
\left({\delta T \over T}\right)_j^2  \right> } } .
\label{e-cdopp}
\end{equation}

In a case where two corresponding pixels have orthogonal 
line-of-sight vectors, the two temperature values are unrelated 
(except if the three-dimensional 
centre of the pixel happens to be at the centre 
of a spherically symmetric gravitational potential),
so a noise value would be contributed to $S$ [eq.~\ref{e-corr}],
but nothing would be contributed to $\Sdopp.$ 

In the case of corresponding pixels whose lines-of-sight are parallel
and equal, contributions would be made both to $S$ 
and to $\Sdopp.$ 
In the case of corresponding pixels whose lines-of-sight are parallel
but opposite, a negative value (or anti-correlation) would be added to $S,$ 
but a positive value would be added to $\Sdopp.$ 

For these reasons, if the Doppler effect (or another directional 
effect) were present (even partially), then
the value of $S$ would be low and noisy in the presence of genuinely
matched circles, to the extent to which the effect is present.
On the other hand, $\Sdopp$ would have a high value, although
less than the ideal value of unity as would be expected for $S$
in the case of locally isotropic emission and a zero
intrinsic (non-topological) auto-correlation function.

\nice \tasdswght ::::

Although $\Sdopp$ is calculated here out of interest, 
it is most likely to be useful in application to 
future measurements by the Planck and MAP satellites.

\subsection{OBSERVATIONS} \label{s-cobe}

The COBE DMR observations and results are discussed by, for example,
\citet{Bennett94}, in which a spherical harmonical analysis 
of the two year results is presented.

In this paper, the results of four years' data 
are used, made available as `DMR Analysed Science
Data Sets' (hereafter, ASDS)
by the COBE team\footnote{At {\em 
http://www.gsfc.nasa.gov/astro/cobe/cobe\_home.html}.},
corrected for
galactic emission either by the `subtraction' or `combination'
techniques of removing synchrotron, dust and free-free emission.
(Hereafter, the `subtracted' and the `combined' maps respectively.) 
The subtracted maps have the higher signal-to-noise ratio of the two.
The weights for the three frequencies used to obtain the two
maps 
(essentially those of \citealt{Bennett94})
are listed in Table~\ref{t-asdswght}.

Because of galactic contamination, data between galactic latitudes
of $-20\deg$ and $+20\deg$ are not considered.

Analysis of the significance of individual fluctuations (as opposed
to their statistical significance) was carried out by 
\citet{CaySm95}. These authors noted (\SS5 of \citeauthor{CaySm95}) 
that some regions outside
of the above galactic cut could also be galactic contaminants.
The possibility that the second, third and seventh hot spots 
listed in Table 2B of  \citeauthor{CaySm95} could be galactic
contaminants due to the Ophiuchus complex and the Orion complex
is considered here.

Since the data set is oversampled, a smoothing by a Gaussian
of minimum full width half maximum (FWHM) of $10\deg$ is necessary
to extract a significant signal. In fact, as is shown below, 
the result is stronger if smoothing with a FWHM of $20\deg$
is considered acceptable.

More recent analyses of galactic contamination of the DMR data 
have been discussed extensively by \citet{Kogut96a,Kogut96b}.
Based on these discussions, 
alternative ways of combining the data from the six DMR antennae
in order to further minimise galactic contamination relative to
the ASDS maps provided by the COBE team, e.g. using a weighted average
of the 53 and 90 GHz maps plus a `custom' Galaxy cut,
would be recommended for further applications of
the circles principle using COBE data.




\section{RESULTS} \label{s-results}

\subsection{Difference parameters: $\sigma$ and $d$}
\nice \tresults ::::

The values of the parameters representing pixel pair 
differences and correlations are listed in Table~\ref{t-results}.
Depending on whether the `subtracted' ASDS or the `combined'
ASDS is used, and on whether smoothing is at the $10\deg$ or
the $20\deg$ scale, the distributions of temperature differences
in corresponding pixel pairs are clearly wider than the $\sigma=1$
distribution expected if the temperatures were intrinsically 
identical and the differences were only due to random measurement
error.

Formally, a Kolmogorov-Smirnov test on the list of the normalised 
temperature
differences used in calculating $\sigma$ and $d$ rejects the 
possibility that 
the distributions of these differences are drawn from a Gaussian 
distribution of mean zero and standard deviation unity by 
at least 94\%. (The weakest rejection in Table~\ref{t-results} 
is for the `combined' data set with $10\deg$ smoothing; the
other rejections are stronger.) 
This is the case whether or not the 
$N\sim1000$ 
pixel pairs 
are considered to be independent, or whether 
only $\sim 150$ of these pairs (half the number of independent pixels) 
are considered to be independent. 
In either case, the large number of realisations of the 
difference distribution results in a precise value of the
dispersion.

This result is only valid if the COBE team's analysis of the
random errors is correct, and if contributions from systematic
errors are negligible. For example, if systematic errors 
or underestimates of the random errors caused the total
uncertainties per pixel to be twice as high as the random
uncertainties presented in the data set, then only the
$20\deg$ smoothed `subtracted' data set would provide
a significant contradiction.

\nice \fcircles ::::

\nice \faitoff ::::

\subsection{Matched circles}

What are the actual values of the temperature fluctuations along the
identified circles? Some examples of
these are plotted in Fig.~\ref{f-circles} and the positions
of two large pairs of circles are shown in Fig.~\ref{f-aitoff}.

From this figure the reader can judge what typical fractions 
of the sky are available for comparison of matching circles, 
given the galactic cut of $\pm20\deg$ latitude and whether 
any systematic effects might be present.

Some of the pairs of circles look impressively similar.

Since the data is smoothed to $20\deg \sim 2000${\hMpc}, 
in the case of pairs of circles for which $\Delta < 2000${\hMpc},
it is clearly possible that some identity between corresponding
circles can be induced just by the smoothing. For example,
the temperature fluctuations around the 
two circles in the panel for $(i,j)=$ $(0,2)$
look very similar, but this is explainable by smoothing as much
as by physical identity.

In the panels for $(-6,3),$ $(-5,2),$ $(-4,1)$ and $(-4,2)$
there are peaks which appear to coincide, though the overall
slopes vary. In fact, these appear as much stronger, coinciding
peaks in the `subtracted' map. This suggests an explanation
as a galactic feature.
In other panels, such as $(-6,4),$ $(-6,5)$ $(-6,6)$ and 
$(-5,5)$ there are strong peaks which are highly in conflict 
between the two circles.

In the latter three cases, it could be argued that the 
temperature differences cannot be used to refute the $T^2$
candidate under consideration, because there could be
contamination from the galactic centre, which may not
have been subtracted perfectly, or may simply have left
residual noise even if well subtracted. A large part 
of the central conflicting peaks in these three panels is
indeed removed if an exclusion zone of $60\deg$ around
the galactic centre is applied.

Moreover, one could also remove areas possibly contaminated 
by the Ophiuchus complex and the Orion complex, as 
suggested by \citet{CaySm95}.

However, as can be seen statistically from the final three
lines of Table~\ref{t-results}, removal of areas possibly
contaminated by the galactic centre and by the Ophiuchus and
Orion complexes removes more well-matched parts of circles
than ill-matched parts, with $\sigma$ increasing from $2.1$ 
to $2.4$ (the $\Omega_0=1, \lambda_0=0$ case is the one illustrated).
For example, the central segments of panels for $(i,j)=$
$(4,6)$ and $(5,5)$ appear consistent within about $1\sigma,$ 
but both lie within $60\deg$ of the galactic centre.

The positions of the Ophiuchus and Orion contaminations used
are those listed as the second, third and seventh lines
of Table 2B of \citet{CaySm95}. The radii of exclusion adopted 
are $57\deg,$ $9\deg$ and $29\deg$ respectively
(based on the pixel numbers of the same table).

Examples of matching segments within $57\deg$ of the first
Ophiuchus component (and not within $60\deg$ of the galactic
centre) are the right-most segments in panels 
$(-3,3)$ and $(-1,6).$
Removal of these possible contaminating segments removes some
conflicting segments, but also some consistent segments.

Finally, if one considers the panels which are least affected
by the galactic cut, in particular panels $(-2,2),$ $(-1,1),$ 
$(5,6),$ $(6,5)$ and $(6,6),$ these seem to show little
disagreement, apart from the right-hand end
of the latter panels. 
Of these, the panel $(-1,1)$ is for 
a pair of circles only separated by $\Delta = 1327${\hMpc}, 
so the agreement can be explained by smoothing rather than 
by topology.

The right-hand end $(5,6),$ $(6,5)$ and $(6,6)$ can be removed
to improve the agreement by arguing that this is
contamination by the Ophiuchus complex, as before.
However, this removal, and also removal 
of the galactic centre (as before) remove parts of the
curves which do agree. If one takes the remaining parts of
the curves, then there is still a broad agreement, for example
if the data is further smoothed on a larger scale. 

But in that case, there 
is the possibility that what matches is merely a quadrupole
or dipole component which has been imperfectly subtracted.
The dipole subtraction for the ASDS's is quoted to be good 
but not perfect, so not only would a different statistic need
to be calculated in order to measure the matching in such 
large scale gradients, but the dipole and quadrupole subtraction
would have to be done precisely enough that this would be 
meaningful.

\nice \fslong ::::

\nice \fdlong ::::

\nice \fclong ::::

\nice \fplong ::::

\section{DISCUSSION} \label{s-disc}

\subsection{The $S$ statistic of \protect\cite{Corn98b}}

If the error estimates are correct, then the statistics
above are sufficient to rule out the present $T^2$ hypothesis.

How successful is the correlation statistic $S$ of \citet{Corn98b}
in refuting the candidate? 

The values of $S$ shown in Table~\ref{t-results} are very low.
Compared with Fig.~2 of \citeauthor{Corn98b}, these
values are quite surprisingly low. The details of the
model for the figure are likely to differ
somewhat from the present model, and the difference in using
simulations versus observational data (in the present case)
could explain some of the difference.

A value close to unity would 
be expected in an ideal case of matching circle pairs,
and a value of zero for non-matching pairs, if the 
two-point auto-correlation function of the temperature
fluctuations were zero. In reality, 
there is an intrinsic non-zero auto-correlation function, 
so a value showing this weak 
a correlation is unexpected. 

If standard arguments implying the relative weakness of
the Doppler effect are invalid, 
could this be due to a dominant contribution by the 
Doppler effect, which would cause genuinely anti-correlated
temperature values to cancel when evaluating the $S$ statistic?
The parameter $\Sdopp$
does not support this. Its value is smaller in absolute value 
than that of $S$, presumably because
typical values of $\nhati {\boldmath .} \nhatj$ are smaller
than unity and there is no underlying anti-correlation.

Figures~\ref{f-slong}-\ref{f-plong} show that the small value of $S$
is just a coincidence. 
These plots
show equivalent hypotheses to that based on the three clusters, apart
from a rotation in galactic longitude.
This avoids any systematic effects due to the
galactic cut or to other functions of galactic latitude. 

Assuming that the hypothesis is false, the ranges of values of the 
statistics for different longitudes show typical
values of the parameters for the $T^2$ models equivalent to the
hypothesised one (equivalent apart from the lack of observational 
motivation in the rotated models).

Fig.~\ref{f-clong} shows why the low value of $S$ 
for the \citet{RE97} candidate is just a coincidence. The 
position of the candidate just happens to lie at a longitude
where both $d$ and $S$ have low (absolute) values. The longitudinal 
dependence of $d$
clearly has a sinusoidal-like component, which could be explained
by some large scale feature in the temperature map.
The longitudinal dependence of $S$ shows more complexity, 
and provides a distribution of maximum values about a factor
of two to four times 
lower than those of `$\Smax$' in \citeauthor{Corn98b}'s
simulation.

\subsection{The $\Sdopp$ statistic}

The shape of the $\Sdopp$ distribution (Fig.~\ref{f-plong}) 
is quite different 
from that of the $S$ distribution (Fig.~\ref{f-clong}). 
Typical values of $\Sdopp$ are roughly half that of the 
typical $S$ values.
An order of magnitude estimate to explain this is that
(very) roughly half of the pairs of matched pixel vectors 
are roughly orthogonal, and so contribute very little to 
$\Sdopp.$ 

Some features in the $S$ distribution, 
e.g. several local maxima and the global minimum
in Fig.~\ref{f-clong}, can be identified with similar 
features in Fig.~\ref{f-plong}. These can be interpreted as 
features due to large matched circles
for which the matched pixel vectors are roughly parallel.

Apart from the `missing' contribution of orthogonal vectors
in Fig.~\ref{f-plong}, the differences between the two figures
also come from roughly opposite vector pairs, i.e. from 
pairs of small circles for large $i$ and $j$ values, where 
the `second observer' is close to $2\rSLS$ from the first
observer.

This demonstrates an approach which will be particularly
useful for the study of MAP and Planck Surveyor data.
The statistic $\Sdopp$ as defined in eq.~(\ref{e-cdopp}) 
will provide a simple method of searching for 
roughly opposite, small identified circles in which 
a Doppler component is strongly expected to make a significant
contribution. 

If a candidate 3-manifold is detected in a new CMB map, then 
comparison of `false' hypotheses rotated in galactic longitude such as 
here will provide a simple way of checking the significance
of the detection, avoiding the need for simulations.

\subsection{Comparison of COBE data to arbitrary 
$T^2$ models via simulated temperature fluctuations}

In the absence of specific candidates for torus models, 
several authors have previously compared COBE data to
sub-classes of these models, but using the perturbation simulational
approach which requires assumptions on perturbation
statistics as described above (\SS\ref{s-twodim}).
The closest work among these to the present study is 
that of \citet{dOSS96}, who considered $T^1$ and $T^2$ 
models.

Because of the presence of one (or two) long axis(es),
the `large-scale cutoff in Fourier mode' argument for $T^3$
models, which can be valid for the assumptions of Gaussian
random amplitudes in the Fourier
modes, does not apply in these cases.

These authors defined a statistic, $S_0,$ which is
similar to $\sigma,$ 
but which includes multiply imaged pixel pairs along 
with many non-multiply imaged pixel pairs.
The advantage of this is that 
the statistic is a minimum value, for which  
the orientation of the fundamental polyhedron is not assumed. 
The disadvantage
is that non-multiply imaged pixel pairs are indiscriminately 
mixed with multiply imaged pixel pairs. 
The points (a), (b), (d), (e) and (f) (\SS\ref{s-disadvant}) also
apply to this study.

This enabled simulations, based on assumptions regarding
the power spectrum of density perturbations,
to be made in a parameter space comprising only the lengths of
the fundamental polyhedron and two random variables for
the Fourier modes, 
but not the orientation of the fundamental polyhedron. The
statistic could therefore be calculated within practical
computing capabilities.

Although this $S_0$ parameter is not identical to $\sigma,$
and it includes noise due to non-corresponding pixels, 
the value obtained is $2.59,$ not too much larger than
the value of $\sigma$ which only compares matching pixels.
This suggests that the $S_0$ parameter could still be
useful for application to Planck and MAP data.

\section{Conclusions} \label{s-conc}

The COBE Advanced Science Data Sets (ASDS) have been used to
try to refute the candidate for the global topology
of the Universe defined by supposing that the
three clusters Coma, RX~J1347.5-1145 and CL~09104+4109 
are three topological images of a single cluster and determine
two generators of the 3-manifold, and that the third generator
is large and orthogonal to the first two. The candidate
is a $T^2 \times R$ candidate, for a flat, i.e. 
$\Omega_0+\lambda_0=1$ universe, where $\lambda_0=0$.

Given the assumptions that 
\begin{list}{(\roman{enumi})}{\usecounter{enumi}}
\item the error estimates given in the ASDS's are sufficient 
as good estimates of the total random plus systematic errors, and
\item the temperature fluctuations measured are dominated 
by the na\"{\i}ve Sachs-Wolfe effect which averaged over 
spherical sections results in locally isotropic emission,
\end{list}
then,
\begin{list}{(\roman{enumi})}{\usecounter{enumi}}
\item using either the `subtracted' or the `combined' ASDS, 
\item with smoothing at either $10\deg$ or $20\deg$, and 
\item with or without making cuts for possible contamination 
by the galactic centre and by the Ophiuchus and Orion complexes,
\end{list}
the difference statistics $\sigma$ and $d$ clearly refute
the $T^2$ hypothesis to more than 94\% significance.
(This is estimated by a Kolmogorov-Smirnov test comparing
the list of temperature differences to a Gaussian 
distribution of mean zero and standard deviation unity.)

A galactic cut of $\pm20\deg$ is used for all the above options.

The subtracted set more strongly rejects the hypothesis than
the combined set. Smoothing at $20\deg$ leads to a stronger
rejection than for smoothing at $10\deg.$
Removal of the suspected contaminating regions 
strengthens the rejection, suggesting that these introduce
more spurious well-matched portions of circles than ill-matched
portions. 

This result does not rely on assumptions on the form of 
the fluctuation spectrum or on the distributions of amplitudes
and phases of the spectrum.


The validity of assumption (i) will be tested by the increasingly
numerous small angle measurements and by the Planck and MAP 
missions. 

In the case of a non-zero cosmological constant ($\lambda_0=0,$ 
e.g. \citealt{FYTY90}), 
the $T^2$ candidate has not yet been ruled out, since 
a modelling of the integrated Sachs-Wolfe (ISW) effect would be
required. 
In this case, assumption (ii) would be invalid to 
the extent that the ISW contributes
to the COBE CMB (\SS7.3, \citealt*{WSS94}). 
If the magnitude of this component is of order of the estimated
noise, then this would be equivalent to assumption (i) failing due
to systematic error. 

It is interesting to note that the matched circles principle 
provides a relatively clean and optimal method of refuting
specific candidates for global topology. For data of the 
COBE quality, it is clear that there is strong
sensitivity to the validity of the noise estimates. Since this
is the case for the matched circles method, it must be an even
stronger caveat for the perturbation simulation methods, which have mostly
been applied to $T^3$ models for which $L_x=L_y=L_z,$ for
which the generators have been assumed to be mutually orthogonal,
and for which simulations of the fluctuation spectra were required.

Since the effectiveness of this new method of applying the identified
circles principle in order to directly refute a 3-manifold candidate
(apart from the caveats) has been shown, future work would be to use
the same method to refute certain classes of flat (or hyperbolic)
multiply connected universe 3-manifolds with COBE data. This would
complement previous work, which relied on assumptions on the
perturbation statistics which are probably inconsistent with the
hypotheses which were tested.


\section*{Acknowledgments}

Helpful discussions and encouragement from 
Tarun Souradeep, Helio Fagundes, Ubi Wichoski, Thanu Padmanabhan and 
Varun Sahni and useful comments from an anonymous referee 
were greatly appreciated. 
Use was made of the COBE datasets 
{\em (http://www.gsfc.nasa.gov/astro/cobe/cobe\_home.html)}
which were developed by NASA's Goddard Space Flight Center 
under the guidance of the COBE Science Working
Group and were provided by the NSSDC. 

\subm \clearpage ::::

\end{document}